# CORMEN: Coding-Aware Opportunistic Routing in Wireless Mess Network

Jeherul Islam and Dr. P K Singh

**Abstract**—These Network Coding improves the network operation beyond the traditional routing or store-and-forward, by mixing of data stream within a network. Network coding techniques explicitly minimizes the total no of transmission in wireless network. The Coding-aware routing maximizes the coding opportunity by finding the coding possible path for every packet in the network. Here we propose CORMEN: a new coding-aware routing mechanism based on opportunistic routing. In CORMEN, every node independently can take the decision whether to code packets or not and forwarding of packets is based on the coding opportunity available.

—————————— ◆ ——————————

## 1  INTRODUCTION

Network Coding [1] has emerged as promising method to increase the network throughput and reliability. Network Coding breaks the conventional store and forward approach of transmission in packet based network. Network coding offers the flexibility of encoding different packets at the intermediate node and transmits it in a single transmission. The receiver node can decode it and extract the specific packet destined for it. The basic idea of network coding is illustrated in Fig. 1 where nodes A, B and C share the common wireless medium. Consider a scenario where nodes A and node C have to exchange some information. Due to the channel constraints only one node can transmit at any given time. The traditional way of accomplishing this information exchange is as follows. Node A sends its packet p1 to the relay node B. The node B forwards this packet to node C. Similarly, node C sends its packet p2 to node B which in turn forwards it to node A as shown in the Figure 1(a). This involves a total of four transmissions.

Now, consider the scenario where network coding is applied to reduce the number of transmissions. The node A and node C transmit packets to central node B sequentially (two transmissions). The node B, instead of directly forwarding each packet to its destination, XOR's the two packets and broadcasts the result as a single packet in the shared medium as shown in the Figure 1(b). Both nodes A and C know their own packet (p1 and p2, respectively) that originated from them. They can therefore retrieve the unknown packet by XORing the known packet with broadcast packet. For example, node A on receiving (p1⊕ p2) performs the operation p1⊕ (p1⊕ p2) to obtain p2. Similarly, node C retrieves packet p1. This entire process takes three transmissions as opposed to four transmissions as is the cse in traditional routing.

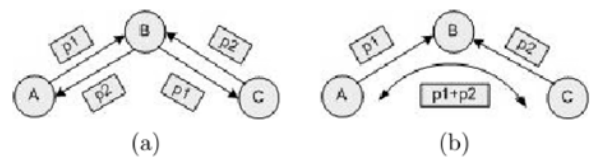

Fig 1. (a) No Coding (b) Network Coding

Opportunistic Routing differs from traditional routing. It exploits the broadcast nature of wireless medium and differs the route selection after packet transmission. Each packet holders chooses a subset of its neighboring nodes as a potential next hope and among those which one have the best node to the destination will be selected as next sender. The main benefits of opportunistic routing is that it can combine multiple weak link into strong link and reduces the retransmission rate especially in wireless medium

In this paper, we are presenting a new routing algorithm which is based on opportunistic routing as well as network coding. The main idea of this algorithm to find a path with minimum delay to the destination which is coding-aware.

## 2  RELATED WORK

The notion of network coding was first proposed by Ahlswede et. al. [1]. They proofed it network utilization can be equal to Max-Flow Min-Cut theory in the context of

————————————————

- *Jeherul Islam is with the ABV-Indian Institute of Information Technology and Management, Gwalior, India.*
- *Dr. P K Singh is with the ABV-Indian Institute of Information Technology and Management, Gwalior, India.*

multicast communication. The Fragouli et. al. [2] provides the brief introduction, benefits and applications of network coding in different areas. The Fragouli et. al. [3] describes the various problems of traditional wireless network and explores to solve these problems by network coding. The Li et. al. [4] discuss Linear Coding, which consider a block of data as a vector over a certain base field and allows a node to apply a linear transformation to vector before passing to it on. They develop a greedy algorithm for code construction. For wireless networks the Katti et. al. [5] proposes COPE, the first practical XOR coding system and demonstrates the throughput gain via implementation and measurement. In COPE every node is in promiscuous mode and overhears the transmission of all other nodes. It uses reception report in their neighborhoods to let the neighbors learn about the packets they currently have. However routing in COPE is independent of coding opportunity. Sengupta at. el. [6] theoretically showed that further coding gain as well as throughput can be achieved if routing is coding dependent.

COPE uses expected transmission count metric (ETX) [7] as its routing function. The ETX [7] metric incorporates the effects of link loss ratios, asymmetry in the loss ratios between the two directions of each link, and interference among the successive links of a path. The ETX of link measures by expected number of transmissions and re-transmissions required by to reliably send a packet across the link. Let $d_f$ be the forward delivery ratio, is the measured probability that a data packets successfully arrives at the recipient and $d_r$ the reverse delivery ratio, is the probability that ACK packet successfully received. The expected probability that a transmission is successfully received and acknowledged is $d_f \times d_r$. Each node measures the delivery rate of its link to and from its neighbor by broadcasting one probe packet every second and counting the number of probes received in the last 10 seconds. The ETX metric is calculated by ETX = $1 / (d_f \times d_r)$. The ETX of a route is sum of the ETX for each link in the route. Opportunistic routing is introduced in [8, 9] The Extremely Opportunistic routing protocol ExOR [8] introduced the concept of being opportunistic when the wireless link are lossy. However it requires lots of coordination among the nodes.

Coding-aware routing was proposed in [6, 10, 11] In [6] authors provide a theoretical framework for investigating the relation between the coding and routing. A localized coding-aware routing algorithm was proposed in [11]. MORE [12] coding-aware routing in Intra-Flow network coding. DCAR [10] is a distributed approach to coding-aware routing. But it chooses a pre-established path i.e. route is selected before sending of packets. But the coding opportunity is depend on traffic flows and at initial stage there is no way to predict the traffic pattern in most wireless networks. CORMEN dynamically select the path in a node-by-node manner during packet forwarding phase.

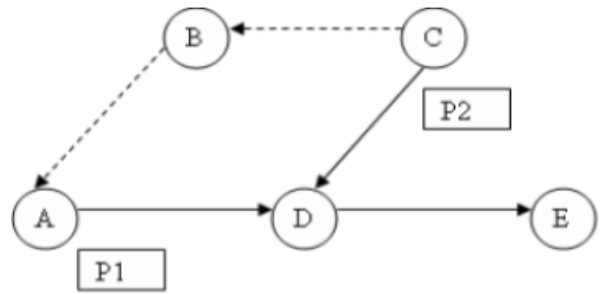

As a result it maximizes the coding opportunity as well as overall network throughput.

## 3 PROPOSED ROUTING ALGORITHM

The routing algorithm in network coding face challenges in finding route which have the highest coding opportunity. The main problem is to determine the exact coding capability at each node in the network. This allows identification of the regions in the network that are more suitable to deploy network coding. This information then can be used in the design of the routing scheme to optimize the network performance. The goal of coding-aware routing is to find out the coding opportunity while forwarding the packet to shortest coding possible path. Design of this algorithms is based on opportunistic routing and the routing metric is depend upon the coding opportunity arises in the intermediate node.

Fig: 2

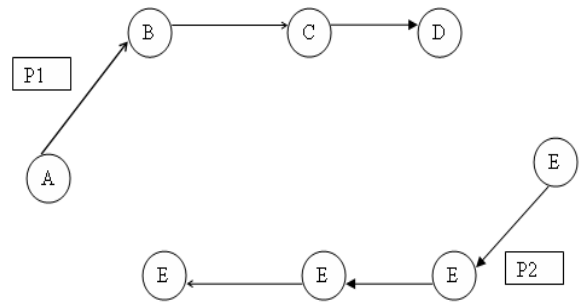

Fig: 3

### 3.1 Design Issues
#### 3.1.1 Routing Metric Selection:

The routing metric must be aware of coding opportunity available at intermediate node. It must choose the best coding possible path. It should not be like that it only chooses the path which has coding opportunity but the route is very long to the destination as well as chooses the shortest path ignoring the coding opportunity.

For Example: consider the case in Fig 2. In this scenario there are two packets P1 and P2 .Packet P1 has source A and destination E. P2 has source C and destination A. If we consider the all links have equal good quality for Packet P2 have two paths CBA and CDA to reach the destination. But P2 shouldn't choose the path CBA as A path have the coding opportunity at node D.

Now consider the second scenario in Fig 3. Packet P1 has source A and destination D. P2 has source E and destination A. Here for packet P2 the EDCBAH path is coding possible but this path is much longer than the non-coding path EFGH. In this case the non-coding path has less delay and more throughputs.

So, the routing metric should select the path in such a way that it not only select the coding possible path but also select the shorter coding impossible path which have higher throughput.

### 3.1.2 Forwarding Node Selection:

Forwarding nodes are the nodes that will accept the packet broadcasted by the sender and among those nodes one will further forward the packet. Forwarding nodes should be selected in advance. This is necessary because the numbers of duplicate transmission and co-ordination overheads tend to increase with number of forwarding nodes. The information of the forwarding nodes store in each packet.

### 3.1.3 Duplicate Transmission Avoidance:

When a sender sends a packet multiple nodes hear the packet. So it must ensure that only the best forwarding node will send the packet to the destination, others will drop the packet.

### 3.1.4 Coding Correctness:

The throughput of the algorithm mainly depend upon the how efficiently and correctly an intermediate node code packets. A node only codes those packets which are decodable at the destination. One most important thing is that the intermediate node must take the coding decision independently without doing any interaction with the neighboring nodes.

## 3.2 Design Decision

### 3.2.1 Forwarding Node Selection:

A sender selects the forwarding nodes for every packet individually. Selection of the forwarding nodes must obey following conditions.

- The ETX between the sender and forwarding nodes must be below threshold limit to ensure of good quality link.
- The forwarding nodes must be adjacent to the shortest path to avoid diverging of path. So each forwarding nodes is close to at least one node on the shortest path.
- The ETX of link between any pair of forwarding nodes within a threshold. This ensures that the forwarding nodes have good connectivity among themselves and to the nodes on the shortest path.

### 3.2.2 Shortest Path Selection:

The shortest path is selected between the source and destination in terms of ETX [5] Metric. This shortest path may or may not be the actual path of a packet. But actual path whether it is coding possible or not must be near to the shortest path. This is required not to diverge the path and avoiding unnecessary duplicate transmission.

### 3.2.3 Coding of Packets:

The overall throughput of the algorithm is depending upon the how efficiently and correctly the node codes the packets. To ensure that the intermediate node code only the proper packets i.e. packets which are decodable at the next hop, it must check the following conditions before coding packets.

Let

   F(c,p) = List of Forwarding nodes for packet 'p' at node 'c'.
   T(c,p) = Node traveled by packet 'p' to reach node 'c'.
   O(c,p)} = List of overhearing node of packet 'p' as it reaches to node 'c'. (**Overhearing Nodes** are the list of nodes who listen and stores the packet when a sender sends it . It is actually the sender's Forwarding Nodes.).

Two packets p1 and p2 will be coded at a node 'c' if anyone of the following conditions holds.

- **Condition 1**: There exist a node a € T(c, p1) such that a € F (c, p2) and there exists a node b € T(c, p2) such that b F(c, p1).
- **Condition 2:** There exists a node a € O(c, p1) such that a € F(c, p2) and there exists a node b € O(c, p2) such that b € F(c, p1).
- **Condition 3**: There exists a node a € O(c, p1) such that a € T(c, p2) and there exists a node b € O(c, p2) such that b € T(c, p1).

### 3.2.4 Setting Forwarding Timer:

Forwarding timer is the most important aspect of CORMEN. The timer is used to avoid duplicate transmission, but it affects the overall throughput of the algorithm. After receiving a packet every node sets a timer for each packet. The value of the timer is depend upon the coding opportunity at that particular node at a particular instant as well as the best route in terms of ETX[7] to the source and destination.

Let 'n' be the no of packets that can be coded at the node and 'i' is the order of the node in the forwarding node set. Then the forwarding timer $t_f$ will be

$$t_f = i/n^2 * ETX_{destination}$$

### 3.2.5 Forwarding of Packets:

Every node maintain two different output queue one is for sending packet originated from the node itself and other is for forwarding of packet received from other node. CORMEN applies two different methods because packets at the source node need not to be coded. So checking coding opportunity is irrevrent. While packets received from other node we must check coding options.

When a source is sending a packet, it simply first select the shortest path to the destination in terms of ETX. It also select list of forwarding nodes and reorder them as per ETX value. It embeds this information on the packet itself. Then it sends the packet as in traditional network and store the packet in a packet pool for some amount time.

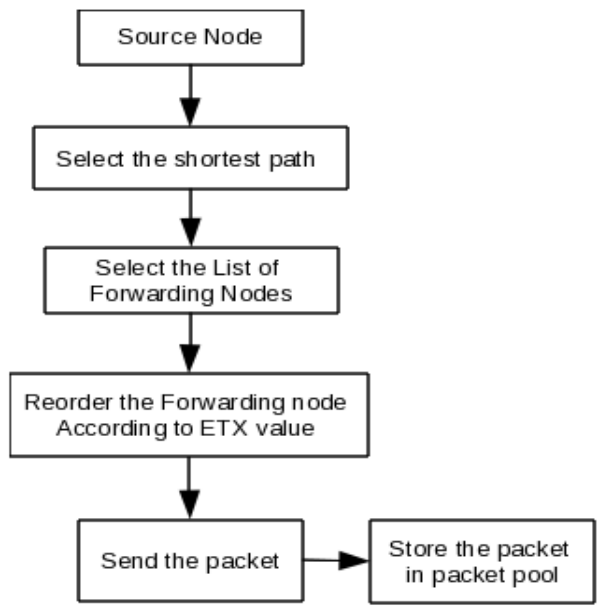

Fig 4 : Flowchart: Source node sending a packet

When an intermediate node receives a packet it first checks whether it is a coded packet or not. If it is coded packet it checks whether it is one of the recipient by seeing the recipient list. If it is not in the recipient list it discards the packet. By doing this it will reduce the unnecessary computation overhead in the node. If this node is in the recipient list, it decodes the packet and checks if it is the destination node for these packets. For which packet it is destination node it sends the acknowledgment. If it is not the destination node, it en-queue the packet in the output queue and sets forwarding timer for the packets. Now updates the forwarding node set and shortest path to the actual destination. It the waits for a probe packets to receive i.e. if there is some other node which have more coding opportunity then that node have a smaller timer. So it will send the probe packet earlier. If the node receives probe packet it will remove the packet from the queue and store the packet in the packet pool. If the timer expires and it will not receive any probe packet, the node sends a probe packet. It the coded the all possible packets and forward it.

## 4 SIMULATION AND PERFORMANCE EVALUATION

We simulate our proposed routing algorithm in NS2 [13] and compare the result with COPE. We simulate COPE from [5]. For simulation of CORMEN, we use IEEE 802.11 MAC, UDP implementation from standard NS distribu-tion. We keep the physical layer bit rate 1 Mbps and use two-ray ground propagation model. We set the transmitting rate interval to 0.05 seconds. Each node has 250 meters radio range and 550 meters carrier sense radius. Each data packet carries 1000 bytes of payload. The RTS/CTS mechanism of 802.11 is switch off in all our simulations.

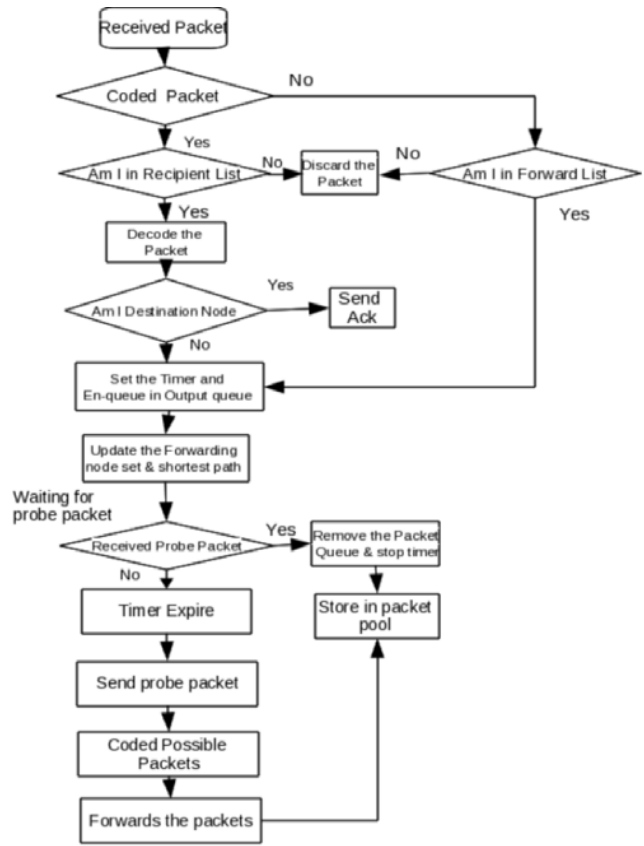

Fig 5: Flowchart: Intermediate Node forwards a packet

### 4.1 Coding Header

A network coding header is constructed to keep necessary information in order to perform coding and decoding properly. We are integrating this header with the IP header. The parameters for these headers are

- NC: It checks whether a packet is native or coded packet. If NC=0, 1 then its native packet. Either if NC = n (n! = 0, 1) then there is n nos of packet coded together.

- FN: It contains the list of forwarding nodes.
- SP: It contains the route information of the default shortest path.

## 4.2 Reference Scenario and Performance Metrics

We consider only grid topologies of different no of nodes for our performance evaluation. Nodes are placed at equal distance of 200 meters apart from each others. All the links between the nodes are considered to be lossless. We take 3 grid structures, 9 nodes of size 3x3, 15 nodes of size 5x3 and 25 nodes of size 5x5. Each grid structure have different traffic load. Following are the performance metric that is used to comparing our algorithm with COPE.

- **Packet Delivery Ratio (PDR):** Packet Delivery Ratio (PDR) is the quotient resulting from the number of successful delivered packets to those generated by sources within the simulation period. It is an important metric which indicates congestion level of the network. PDR measures the protocol performance from loss ratio experienced at the network layer that is affected by factors such as packet size, network load and effects of movement resulting in frequent topological changes. Higher PDR implies that packet loss rate is lower and protocol is more efficient from the perspective of data delivery. A point to note is that late packet received could be deemed useless even with high PDR.

PDR= **(**No of Received packet) / **(**No of transmitted Packet)

- **Average End-to-End Delay:** There are lots of possible delays caused by buffering during route discovery latency, queuing at the interface queue, and delay due to coding of packets at the node, retransmission delays at the MAC and propagation and transfer time. End-to-end delay is the overall time difference from transmitting packets from a source to the arrival of packets at the destination node. Average end-to-end delay can be computed as time difference between every packet sent and received, dividing by the total no of packets received at the destination. This metric describes the packet delivery time: the lower the end-to-end delay the better the application performance.

Average End-toEnd Delay =
(Time $_{packet\ arrived\ @dest}$ − Time $_{packet\ sent\ @\ source}$)/ Total no of Packet Recieved

- **Average Throughput:** Network throughput determine the amount of data successfully received at the destination per unit time. Average network throughput can be calculated by the following

Average Throughput= **(**No of bits received at each node )/ Simulation Time

## 4.3 Result Analysis

In 3x3 grid structure there are total 9 nodes and 7 nos of CBR data flows starting at different simulation time. First data flow was started at 30 seconds of simulation time and others are starting at an interval of 20 seconds. So, as the time increases the load of the network also increases. Our main aim is to see the behavior of the network at light traffic as well as heavy traffic. We observed that the coding opportunity increases in heavy traffic. As a result, both CORMEN and COPE achieve higher throughput with increase of traffic. But as the traffic increase in the network due to coding of packets at node the end-to-end delay also increases. In Fig.5 (a) we showed the throughput of COPE and CORMEN. Graph shows that the CORMEN has higher throughput than COPE. Packet delivery ratio Fig.5 (b) is also slightly better than COPE. Average end-to-end delay is also lesser than COPE. Fig. 5(c) shows this. In 3x5 grid structure total 11 nos of CBR flow and in 5x5 grid structure 14 nos of CBR data flow is there. The result of these scenarios is shown in Fig. 6 and Fig. 7. Here also CORMEN is performing better than the COPE.

## 5 CONCLUSION

In this paper, we propose CORMEN, a coding-aware routing algorithm for wireless network. It searches the possible coding opportunity in the route selection process and forwards the packet through the best route. Coding decision can be taken by each node individually, not much synchronization is required among nodes.

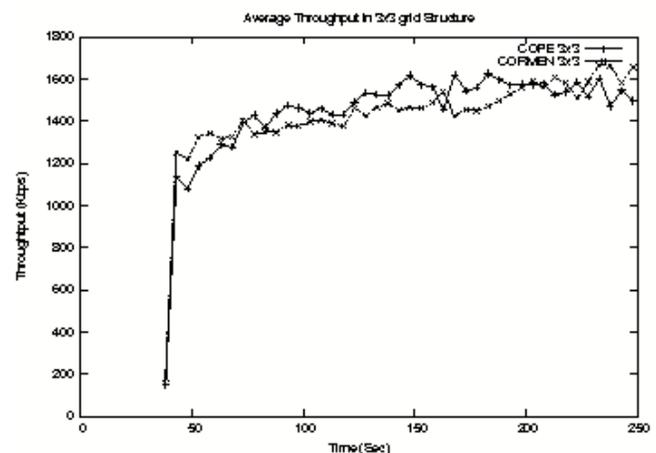

Figure : 5 (a) Throughput 3x3 grid structure

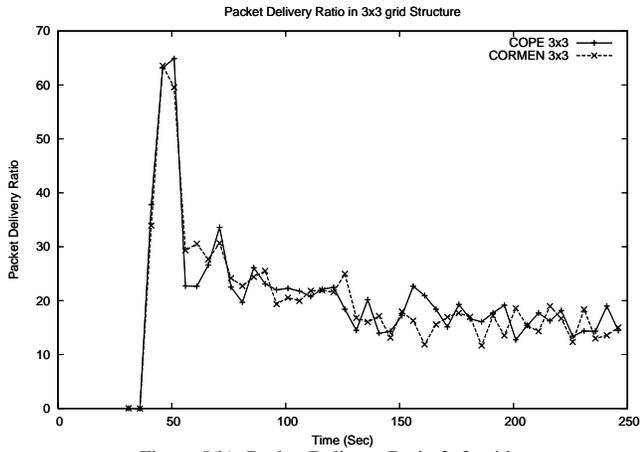

Figure 5(b): Packet Delivery Ratio 3x3 grid

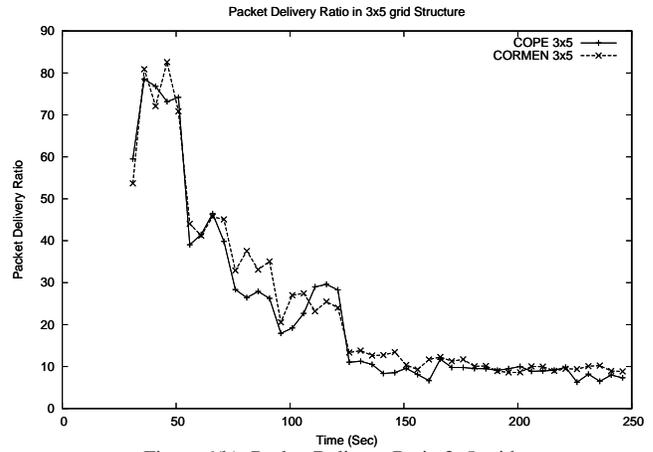

Figure 6(b): Packet Delivery Ratio 3x5 grid

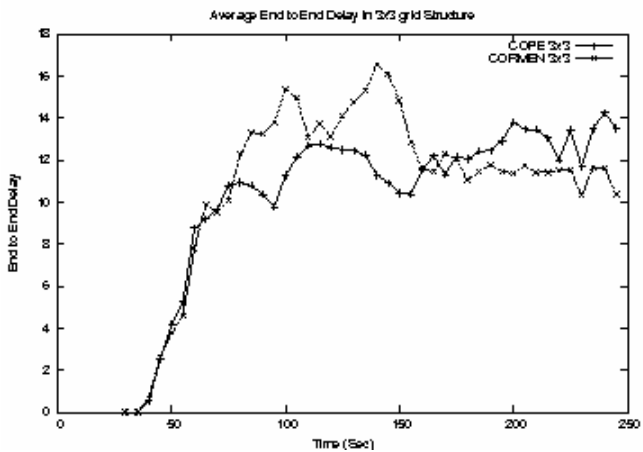

Figure 5(c) End-to-End delay in 3x3 grid

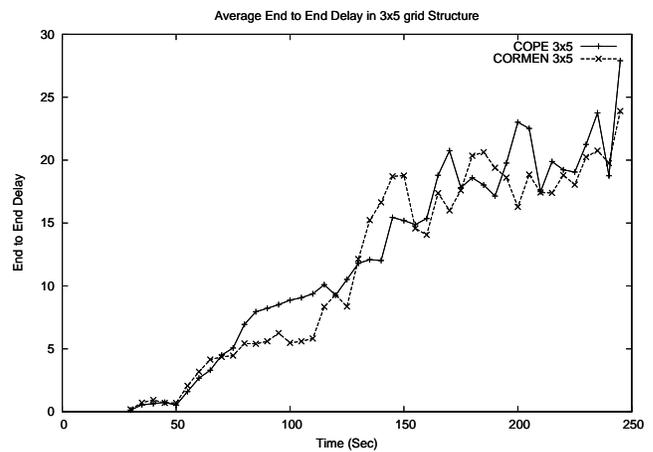

Figure 6(c) End-to-End delay in 3x5 grid

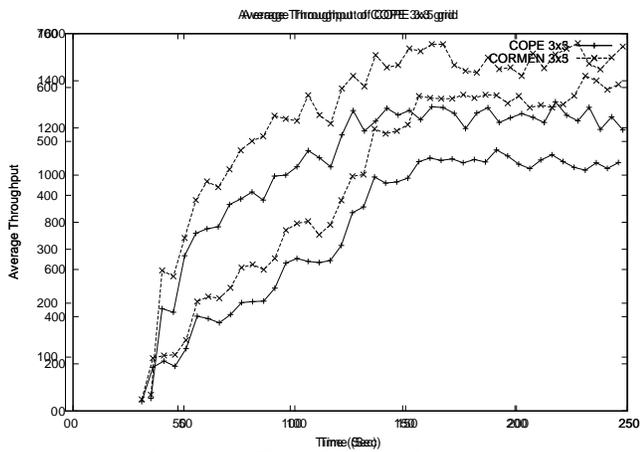

Figure : 6 (a) Throughput 3x5 grid structure

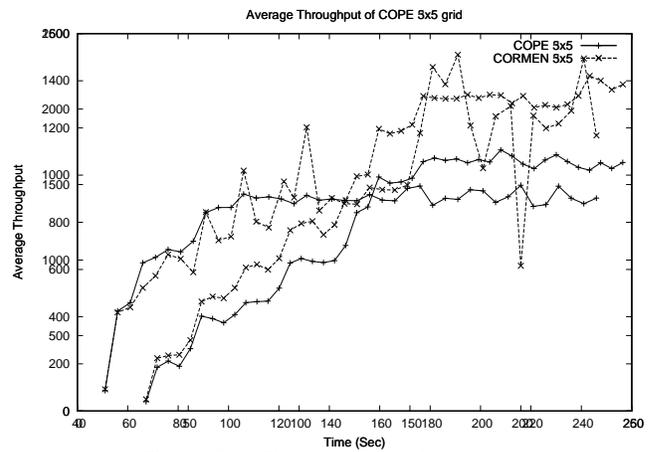

Figure : 7 (a) Throughput 5x5 grid structure

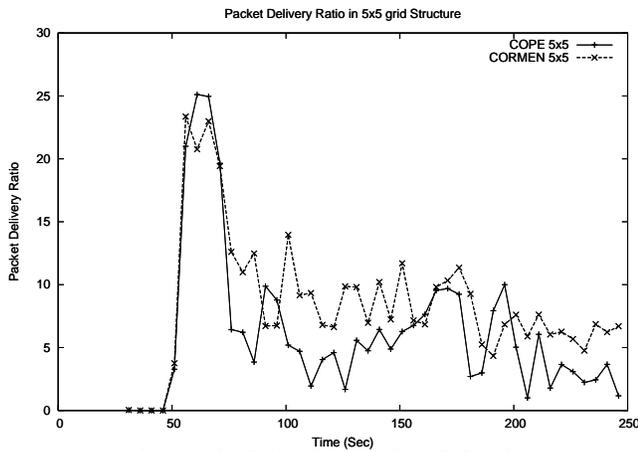

Figure 7(b): Packet Delivery Ratio 5x5 grid

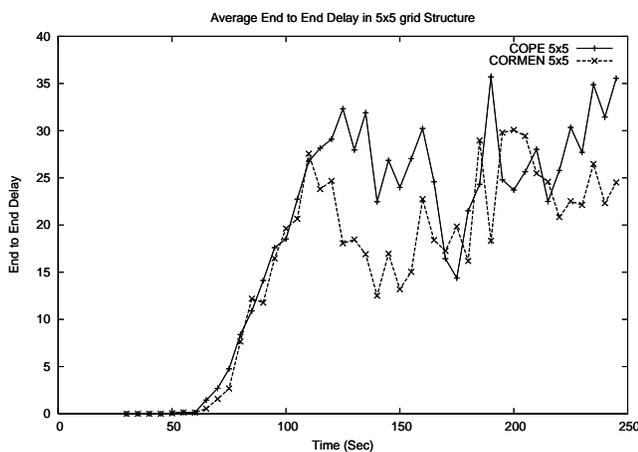

Figure 7(c) End-to-End delay in 5x5 grid

**Jeherul Islam** is a student of Master of Technology at ABV-Indian Institute of Information Technology and Management, Gwalior, India. He received his Bachelor of Engineering degree in CSE form Guwahati University, India. Currently working as a Scientific Officer at Indian Institute of Technology Guwahati, India. His research interest is wireless networking, network coding.

**Dr P K Singh** received his B. Tech. (B.S.) degree in CSE from Kamla Nehru Institute of Technology Sultanpur, India in 1989, and M.Tech. (M.S.) in CIT and Ph.D.degrees from Indian Institute of Technology Kharagpur, India in 1998 and 2008 respectively. He was associated with NIT Jalandhar, India as Lecturer/Senior Lecturer till 1999 and as an Assistant Professor at SLIET Longowal, India till 2001 and as Senior Networking Engineer at IIT Kharagpur, India till 2008. Currently he is associated with ABV-IIITM Gwalior, India as Associate Professor. His current research interest includes computer networks, evolutionary computation, multiobjective optimization and data mining. He has coauthored more than 30 technical publications as book chapters and articles in journal/conference proceedings.